%Paper: hep-th/9409108
%From: zamolod@lpm.univ-montp2.fr (Zamolodchikov Aliocha)
%Date: Mon, 19 Sep 94 19:06:40 +0100

\magnification=1200
\parskip 3pt plus 1pt

\font\tit=cmr10 scaled \magstep2
\font\nom=cmr10 scaled \magstephalf

\def\inbar{\,\vrule height1.5ex width.4pt depth0pt}
\def\IB{\relax{\rm I\kern-.18em B}}
\def\IC{\relax\hbox{$\inbar\kern-.3em{\rm C}$}}
\def\ID{\relax{\rm I\kern-.18em D}}
\def\IE{\relax{\rm I\kern-.18em E}}
\def\IF{\relax{\rm I\kern-.18em F}}
\def\IG{\relax\hbox{$\inbar\kern-.3em{\rm G}$}}
\def\IH{\relax{\rm I\kern-.18em H}}
\def\II{\relax{\rm I\kern-.18em I}}
\def\IK{\relax{\rm I\kern-.18em K}}
\def\IL{\relax{\rm I\kern-.18em L}}
\def\IM{\relax{\rm I\kern-.18em M}}
\def\IN{\relax{\rm I\kern-.18em N}}
\def\IO{\relax\hbox{$\inbar\kern-.3em{\rm O}$}}
\def\IP{\relax{\rm I\kern-.18em P}}
\def\IQ{\relax\hbox{$\inbar\kern-.3em{\rm Q}$}}
\def\IR{\relax{\rm I\kern-.18em R}}
\font\cmss=cmss10 \font\cmsss=cmss10 at 7pt
\def\IZ{\relax\ifmmode\mathchoice
{\hbox{\cmss Z\kern-.4em Z}}{\hbox{\cmss Z\kern-.4em Z}}
{\lower.9pt\hbox{\cmsss Z\kern-.4em Z}}
{\lower1.2pt\hbox{\cmsss Z\kern-.4em Z}}\else{\cmss Z\kern-.4em Z}\fi}
\def\IGa{\relax\hbox{${\rm I}\kern-.18em\Gamma$}}
\def\IPi{\relax\hbox{${\rm I}\kern-.18em\Pi$}}
\def\ITh{\relax\hbox{$\inbar\kern-.3em\Theta$}}
\def\IOm{\relax\hbox{$\inbar\kern-3.00pt\Omega$}}

\hbox{}
\vskip 0.0cm
\line{\hfill\hbox{}}
\vskip 1cm
\centerline{\tit
Painlev\'e III and 2D Polymers
}
\vskip 0.7cm
\centerline{\nom Al.B.Zamolodchikov}
\vskip 0.3cm
\centerline{Laboratoire de Physique Math\'ematique\footnote{*}{\rm
Laboratoire Associ\'e au CNRS URA 768}
}
\centerline{Universit\'e des Sciences et Techniques du Languedoc}
\centerline{Place Eug\`ene Bataillon, 34095 Montpellier Cedex 5}
\vskip 1cm

{\bf Abstract\hfil}

Recently the scaling function of the dilute non-contractible self-avoiding
2D polymer loop on a cylinder was related to the Painlev\'e III transcendent.
Using the perturbation theory, the thermidynamic Bethe ansatz and
numerical calculations we argue a similar relation for the contractible
self-avoiding loop.

\hbox{}
\vskip 1.0cm

{\bf 1. Introduction}

Among the recent studies in the 2D integrable relativistic field theory (RFT)
one of the most impressive results was achieved by P.Fendley and H.Saleur in
ref.[1] (see also related ref.[2]). The authors have succeeded to link the
universal scaling funcition of a single 2D self-avoiding loop winding once
around a cylinder to a particular Painlev\'e III transcendent. After almost
20 years  of the famous two-spin Ising scaling correlation function [3]
standing alone, a non-linear differential equation appears again in 2D RFT
to govern the universal scaling behavior. Being related to an apparently
interacting field theory the result of [1] seems extremely suggestive.
Moreover, it solves exactly an important problem in 2D polymer statistics and
is directly comparable with series expansions, simulations etc. It should
be mentioned that the discovery was preceded by a series of fascinating studies
in the topological and $N=2$ supersymmetric 2D RFT's [4--6]. Although it is
not obvious that it is the $N=2$ SUSY that plays the most important role in the
phenomenon of [1], the physical significance of ref.[4] worth to be better
understood (see ref.[2] in this connection).

The problem of 2D self-avoiding polymers is formulated basically as follows.
Consider some 2D lattice (say the honeycomb one to get rid of any problems
with self-avoiding) and count the continuous self-avoiding paths (closed or
open) through the links of the lattice. In general there are several
disconnected components, each of them being called the polymer (respectively
closed and open). Because of the self-avoiding there are many topologically
different configurations and therefore an innumerable amount of various
counting problems. As a first step one tries to count the closed polymer
configurations taking into account the number of separate polymers. The
configurations can be generated e.g. by the $K$-expansion of the $O(n)$-model
partition function
$$
Z(K,n)={\rm tr}\prod_{<i,j>}\left(1+Ks^a_is^a_j\right)
\eqno(1.1)
$$
where the product is over the nearest neighbour pairs of sites $<i,j>$. The
$n$-component spins $s_i^a$; $a=1,2,\ldots,n$ sit at the sites and are chosen
so that ${\rm tr}s^a=0$, ${\rm tr}s^as^b=\delta^{ab}$ and
${\rm tr}s^as^bs^c=0$. For (1.1)
we have
$$
Z(K,n)=\sum_{{\rm loop}\atop{\rm configuraitons}}
K^{{\rm \# of\ links}}n^{{\rm\# of\ loops}}
\eqno(1.2)
$$
As usual for a homogeneous lattice of large volume $V$ the thermodynamic limit
is supposed to exist
$$
{\cal F}(K,n)=-{1\over V}\lim_{V\to\infty}\log Z(K,n)
\eqno(1.3)
$$
defining the specific free energy ${\cal F}(K,n)$.

What is interesting for the field theory is the statistics of large loop
configurations (i.e. with the number of lattice links in a polymer tending to
infinity) which is believed to be universal, i.e. independent on the lattice
realization. The loops blow up near some critical value $K_c$ where the
statistical quantities (like the free energy) develop singularities
characteristic for the second order phase transition. The phase at $K<K_c$ is
called the dilute polymer one. While $K<K_c$ the correlation length (i.e. the
typical spatial extension of the polymers), which we denote $M^{-1}$ here,
remains finite but diverges as
$$
M^{-1}\sim (K_c-K)^{-(p+1)/4}
\eqno(1.4)
$$
The standard parameter $p$ which appears in the exponent of eq.(1.4) is
related to $n$ as follows
$$
n=2\cos{\pi\over p}
\eqno(1.5)
$$
Near the critical point the observables (correlation functions etc.) apportion
some singular in $K_c-K$ parts, which depend only on the distances and
extents scaled in the units of $M^{-1}$ and bear the rotational (Lorentz)
symmetry characteristic for RFT. Being universal these scaling functions
are objects of the field theory. E.g. the free energy (1.3) itself behaves as
$$
{\cal F}(K,n)=\tilde{\cal F}(K,n)+M^2 f(n)
\eqno(1.6)
$$
where $\tilde{\cal F}$ is some non-universal background regular in $K-K_c$
which is typically of no interest for the field theory. Contrary, the second
piece in (1.6) is contributed by the asymptotically large polymer loops and
describes their universal statistics independent on how the polymers are
arranged on the lattice level. For the 2D polymer problem $f(n)$ is known
exactly
$$
f(n)=-{1\over 4}\tan{\pi p\over 2}
\eqno(1.7)
$$
This result comes both from the relation to the sin-Gordon model (see below)
where the exact vacuum energy was found by different methods [7--9] and from
a formal analytic continuation of the thermodynamic Bethe ansatz (TBA)
equations at integer $p$ [10].

{}From eq.(1.2) it is clear that the $n$-expansion
$$
f(n)=\sum_k f_k n^k=-{1\over 4}n-{1\over 4\pi}n^2-{1\over 4\pi^2}
\left(1+{3\pi^2\over 8}\right)n^3+\ldots
\eqno(1.8)
$$
is in fact the expansion in the number of separate polymer loops. E.g. the
first number $f_1=-1/4$ corrsponds to the (scaling) internal statistical
weight (or activity) of a single isolated loop. In other words $f_1$ is the
specific free energy of a dilute gas of non-interacting contractible polymer
loops. The second two-polymer number $f_2=-1/4\pi$ receives in fact
contributions from two topologically different polymer configurations drawn
in fig.1. The first one (fig.1a) gives the first virial coefficient for the
gas of contractible loops which takes into account their interaction (due to
the self-avoiding) at the two-component cluster level. The second contribution
(fig.1b) is the activity of the structure with two nested loops. Further
numbers in (1.8) include more and more distinct topological contributions. It
is an interesting problem (unsolved to my knowledge) to separate them.

However, in the present note we deal with another settlement of the polymer
problem. In the next section the counting problem of closed polymers on an
infinite cylinder is considered and the corresponding scaling functions are
defined. Then the remarkable result of ref.[1] is quoted which relates
explicitly the one non-contractible polymer scaling function to a particular
Painlev\'e III transcendent. We present also a similar relation for the scaling
function of a single contractible polymer. Sect.3 contains few details and
propositions about the Painlev\'e III transcendents including their
representations as Fredholm determinants as well as the TBA-like
representations [1,2]. In sect.4 we discuss the useful relation between the
cylinder polymer scaling functions and the finite-temperature free energy of
the sin-Gordon model (SG) in the repulsive regime $\beta^2>4\pi$. In
particular this includes the restricted sin-Gordon models (RSG), i.e., the
$\Phi_{13}$ perturbed minimal conformal field theories (CFT's). Few first
terms in the perturbative expansions of the SG and RSG finite-temperature
free energy are developped in sect.5. They are to be compared with the
corresponding expansion of the Painlev\'e III. Sect.6 contains the TBA
considerations in the sin-Gordon model near its $N=2$ supersymmetric point
$\beta^2=16\pi/3$. These permit us to derive the TBA-like formulas quoted in
sect.3. A bulk of questions and hints arises in connection with the things
considered. Some of them are listed in sect.7.

\hbox{}
\vskip 0.5cm

{\bf 2. Polymers on a cylinder. The Painlev\'e III}

Let us start again from the lattice level. Imagine a long cylinder made
of say again the honeycomb lattice (fig.2) and denote $L$ the length of the
cylinder and $R$ its circumference. When counting the closed polymer
configurations on this cylinder lattice one can distinguish between two kinds
of polymers, the ``winding loops'' which wind once around the cylinder
(fig.3a) (it is plain that the self-avoiding allows a closed polymer to do it
only once) and the ``non-winding loops'' which don't (fig.3b). We shall count
these two kinds separately defining the partition function as
$$
Z(m,n|K,R,L)=\sum_{{\rm cylinder\ polymer}\atop{\rm configurations}}
K^{\rm \# of\ links}m^{\rm \# of\ winding\ loops}n^{\rm \# of\
non-winding\ loops}
\eqno(2.1)
$$
where $m$ is the weight of a winding polymer and $n$ is that of a non-winding
one. At $L\to\infty$ as usual we introduce the free energy per unit length
$$
{\cal E}(m,n|K,R)=-{1\over L}\lim_{L\to\infty}\log Z(m,n|K,R,L)
\eqno(2.2)
$$

Of course, at $R$ fixed (in the lattice units) ${\cal E}(m,n|K,R)$ shows no
criticality near $K_c$. It does however if $R$ simultaneously goes to
infinity as fast as the correlation length $M^{-1}$ does. Defining the scaling
circumference
$$
t=MR
\eqno(2.3)
$$
we expect that at $t$ fixed and $K\to K_c-0$
$$
{\cal E}(m,n|K,R)=\tilde{\cal E}(m,n|K,R)+MF(m,n|t)
\eqno(2.4)
$$
Here $\tilde{\cal E}(m,n|K,R)$ is non-singular at $K_c$ and in fact for $R$
large enough
$$
\tilde{\cal E}(m,n|K,R)=R\tilde{\cal F}(K,n)
\eqno(2.5)
$$
independently on $m$ (with the same $\tilde{\cal F}(K,n)$ as in eq.(1.6)).
What is interesting for us is the universal scaling function $F(m,n|t)$ which
describes the statistics of large (dilute) polymers on the cylinder. Obviously
at $t\to\infty$ ($R\gg M^{-1}$) the influence of the cylinder geometry
disappears and
$$
F(m,n|t)\sim tf(n)\ ,\ \ \ \ \ \ t\to\infty
\eqno(2.6)
$$
with $f(n)$ defined in the previous section. For what follows it is convenient
to introduce also the special notations
$$
F_+(n|t)=F(n,n|t)
\eqno(2.7)
$$
for the scaling function of equally weighted winding and non-winding loops and
also
$$
F_-(n|t)=F(-n,n|t)
\eqno(2.8)
$$
for that weighted with the opposite signs.

Again the double expansion in $m$ and $n$
$$
\eqalign{
F(m,n|t)&=\sum_{l_1,l_2}F_{l_1,l_2}(t)m^{l_1}n^{l_2}\cr
&=mF_{1,0}(t)+nF_{0,1}(t)+m^2F_{2,0}(t)+mnF_{1,1}(t)+n^2F_{0,2}(t)+\ldots
}\eqno(2.9)
$$
isolates the configurations with fixed numbers $l_1$ of winding loops and $l_2$
of non-winding ones. Each of the linear terms $F_{1,0}(t)$ and $F_{0,1}(t)$
corresponds respectively to the scaling statistical weight (per unit length of
the cylinder) of a single isolated non-contractible (fig.3a) and contractible
(fig.3b) polymer loop. The next quadratic terms in expansion (2.9) contain
different two-loop scaling functions. Namely, $F_{2,0}(t)$ and $F_{1,1}(t)$
are related to the two-polymer clusters drawn in figs.4a and 4b while
$F_{0,2}(t)$ is contributed by two topologically different configurations of
figs.4c and 4d.

The one-loop cylinder scaling functions are of primary interest below so that
we introduce special notaitons for them
$$
\eqalign{
F_n(t)&=F_{1,0}(t)\cr
F_c(t)&=F_{0,1}(t)
}\eqno(2.10)
$$
In connection with the functions (2.7) and (2.8) we shall also consider the
combinations
$$
F_{\pm}(t)=F_c(t)\pm F_n(t)
\eqno(2.11)
$$
The remarkable result of ref.[1] is that
$$
F_n(t)={dU(t)\over dt}
\eqno(2.12)
$$
where $U(t)$ is a particular solution (regular at $t>0$) to the Painlev\'e III
equation
$$
{1\over t}{d\over dt}t{d\over dt}U={1\over 2}\sinh 2U
\eqno(2.13)
$$
This equation admits a one parameter family of regular at $t>0$ solutions (see
e.g.[11]) called the Painlev\'e III transcendents. The special solution in
eq.(2.12) is fixed by the following boundary conditions at $t\to 0$
$$
U(t)=-{1\over 3}\log t-{1\over 2}\log{\kappa\over 4}+O(t^{4/3})
\eqno(2.14)
$$
where we have denoted
$$
\kappa={\Gamma^2(1/3)\over\Gamma^2(2/3)}=3.91392514\ldots
\eqno(2.15)
$$
{}From eqs.(2.13) and (2.14) it is quite easy to see that $U(t)$ is essentially
a regular series
$$
U(t)=-{1\over 4}\log\xi-{1\over 2}\log{\kappa\over 4}+\sum_{n=1}^\infty
U_n\xi^n
\eqno(2.16)
$$
in the variable
$$
\xi=t^{4/3}
\eqno(2.17)
$$
The coefficients $U_n$ can be recursively generated by the equation (2.13)
$$
{16\over 9}\sum_{n=1}^\infty n^2U_n\xi^n=
{\xi\over\kappa}e^{2\sum_{n=1}^\infty U_n\xi^n}-{\xi^2\kappa\over 16}
e^{-2\sum_{n=1}^\infty U_n\xi^n}
\eqno(2.18)
$$
This way one finds
$$
\eqalign{
tF_n(t)&=-{1\over 3}+{4\over 3}\sum_{n=1}^\infty nU_n\xi^n\cr
&=-{1\over 3}+{3\over 4\kappa}\xi-{3(\kappa^3-18)\over 128\kappa^2}\xi^2
+{27(\kappa^3+18)\over 2048\kappa^3}\xi^3-{27(\kappa^6+30\kappa^3-648)\over
131072\kappa^4}\xi^4+\ldots
}\eqno(2.19)
$$
On the other hand the Painlev\'e III transcendent $U(t)$ admits the following
infinite series representation [11,2]
$$
U(t)=\sum_{k=0}^\infty{2\over 2k+1}\int\prod_{i=1}^{2k+1}
{\displaystyle e^{-t\cosh\theta_i}\over\displaystyle\cosh{\theta_i-\theta_{i+1}
\over 2}}{d\theta_i\over 4\pi}
\eqno(2.20)
$$
where it is implied that $\theta_{2k+2}=\theta_1$. The series is
convergent for all $t>0$ while the first terms control the $t\to\infty$
asymptotics. In particular the scaling function $F_n(t)$ falls off
exponentially at $t\to\infty$
$$
F_n(t)=-{1\over\pi}K_1(t)+O\left(e^{-2t}\right)
\eqno(2.21)
$$
where
$$
K_\nu(t)={1\over 2}\int{\displaystyle e^{\nu\theta-t\cosh\theta}}d\theta
\eqno(2.22)
$$
is the modified Bessel function. The non-contractible one-loop function
$F_n(t)$ is plotted in fig.5.

Below we argue that the contractible one-loop scaling function $F_c(t)$
satisfies the following equation
$$
{1\over t}{d\over dt}tF_c(t)=-{1\over 2}\cosh 2U
\eqno(2.23)
$$
where $U(t)$ is the same Painlev\'e III transcendent (2.20) as for the
non-contractible function $F_n(t)$.

Since $U(t)\to 0$ at $t\to\infty$ it follows from eq.(2.23) that $F_c(t)$
has a linear leading asymptotic in this limit
$$
F_c(t)\sim -{t\over 4}\ \ \ \ \ \ \hbox{at}\ \ t\to\infty
\eqno(2.24)
$$
in agreement with (2.6) and the expansion (1.8) of the bulk free energy.
Eq.(2.23) defines $F_c(t)$ up to an item of $C/t$ with an arbitrary constant
$C$. It seems natural to choose $C$ so that the corrections to (2.24) at
$t\to\infty$ would be exponentially small. Then we have
$$
F_c(t)=-{t\over 4}+{1\over\pi^2 t}\int_t^\infty K_0^2(r)rdr+O\left(e^{-4t}
\right)
\eqno(2.25)
$$
In the next section it is argued that in fact $F_c(t)$ develops a series
representation quite similar to eqs.(2.12) and (2.20). Define $V(t)$ as
$$
V(t)={t^2\over 8}+\sum_{k=1}^\infty{2\over 2k}\int\prod_{i=1}^{2k}
{\displaystyle e^{-t\cosh\theta_i}\over\displaystyle\cosh{\theta_i-\theta_{i+1}
\over 2}}{d\theta_i\over 4\pi}
\eqno(2.26)
$$
where again $\theta_{2k+1}=\theta_1$ in each term in the r.h.s. (note that now
we sum up the ``even loops'' instead of the ``odd'' ones in eq.(2.20)). Then
$$
F_c(t)=-{dV(t)\over dt}
\eqno(2.27)
$$
At small $t$ the contractible loop scaling function is again a regular series
in the variable (2.17)
$$
tF_c(t)=c-{4\over 3}\sum_{n=1}^\infty nV_n\xi^n
\eqno(2.28)
$$
The coefficients $V_n$ are readily restored from eq.(2.23)
$$
{16\over 9}\sum_{n=1}^\infty n^2 V_n\xi^n=
{\xi\over\kappa}e^{2\sum_{n=1}^\infty U_n\xi^n}+{\xi^2\kappa\over 16}
e^{-2\sum_{n=1}^\infty U_n\xi^n}
\eqno(2.29)
$$
As it was mentioned before, the constant term $c$ in (2.28) in not fixed by
eq.(2.23). We shall see shortly that the above choice of the integration
constant $C$ in eq.(2.25) corresponds to $c=1/18$. Expansion (2.28) reads
$$
tF_c(t)={1\over 18}-{3\over 4\kappa}\xi-{3(\kappa^3+18)\over 128\kappa^2}\xi^2
+{9(5\kappa^3-54)\over 2048\kappa^3}\xi^3-{27(\kappa^6+6\kappa^3+648)\over
131072\kappa^4}\xi^4+\ldots
\eqno(2.30)
$$
The contractible scaling function $F_c(t)$ is presented in fig.6 (without the
``bulk energy'' term $-t/4$).

For the later use we quote also the short-distance expansions of the
combinations (2.11)
$$
tF_+(t)=-{5\over 18}-{3\kappa\over 64}\xi^2+{9\over 256}\xi^3-
{27(\kappa^3+18)\over 65536\kappa}\xi^4+\ldots
\eqno(2.31)
$$
and
$$
tF_-(t)={7\over 18}-{3\over 2\kappa}\xi-{27\over 32\kappa^2}\xi^2+
{9(\kappa^3-54)\over 1024\kappa^3}\xi^3+{81(\kappa^3-54)\over 16384\kappa^4}
\xi^4+\ldots
\eqno(2.32)
$$
These two functions are plotted in fig.7. They satisfy the equations
$$
{1\over t}{d\over dt}tF_\pm(t)=-{1\over 2}e^{\mp 2U}
\eqno(2.33)
$$

\hbox{}
\vskip 0.5cm

{\bf 3. Painlev\'e III vs. Fredholm theory}

We begin with the second kind Fredholm equation
$$
f(\theta)-\lambda\int_{-\infty}^\infty K(\theta,\theta')f(\theta')d\theta'=
g(\theta)
\eqno(3.1)
$$
where $K(\theta,\theta')=K(\theta',\theta)$ is a symmetric kernel (bounded in
$L_2(-\infty,\infty)$) and $\lambda$ is the spectral parameter. Below we take
the kernel of the following special form
$$
K(\theta,\theta')={\displaystyle e^{-u(\theta)-u(\theta')}\over\displaystyle
\cosh{\theta-\theta'\over 2}}
\eqno(3.2)
$$
with some suitably chosen ``potential'' $u(\theta)$. For the Painlev\'e III
theory $2u(\theta)=t\cosh\theta$ will be relevant. Note, that in terms
of $x=e^\theta$ and substituting $f(\theta)=\exp(\theta/2)\tilde f(x)$,
$g(\theta)=\exp(\theta/2)\tilde g(x)$ eq.(3.1) reads
$$
\tilde f(x)-2\lambda\int_0^\infty {\displaystyle e^{-u(x)-u(x')}\over
x+x'}\tilde f(x')dx'=\tilde g(x)
\eqno(3.3)
$$
The spectral information about (3.1) is contained in the Fredholm determinant
$D(u|\lambda)$, which is entire function of $\lambda$ with zeroes located at
the
eigenvalues $\lambda_a$, $a=1,2,\ldots$ of (3.1). It can be calculated as a
convergent series in $\lambda$
$$
D(u|\lambda)=\sum_{n=0}^\infty{(-\lambda)^n\over n!}D_n(u)
\eqno(3.4)
$$
with
$$
\eqalign{
D_n(u)&=\int_0^\infty\det\left({1\over x_i+x_j}\right)\prod_{i=1}^n
2e^{-2u(x_i)}dx_i\cr
&=\int_0^\infty\prod_{i>j}^n{(x_i-x_j)^2\over(x_i+x_j)^2}\prod_{i=1}^n
e^{-2u(x_i)}{dx_i\over x_i}
}\eqno(3.5)
$$
Denote
$$
-W(u|\lambda)=\log D(u|\lambda)
\eqno(3.6)
$$
For $W(u|\lambda)$ we have the following $\lambda$-series
$$
W(u|\lambda)=\sum_{n=1}^\infty{\lambda^n\over n}W_n(u)
\eqno(3.7)
$$
convergent while $|\lambda^{-1}|>\sup_a|\lambda_a^{-1}|$. Here
$$
W_n(u)=\int\prod_{i=1}^n{\displaystyle e^{-2u(\theta_i)}\over\displaystyle
\cosh{\theta_i-\theta_{i+1}\over 2}}d\theta_i=
\int_0^\infty\prod_{i=1}^n{\displaystyle 2e^{-2u(x_i)}\over\displaystyle
x_i+x_{i+1}}dx_i
\eqno(3.8)
$$
where we imply that $\theta_{n+1}=\theta_1$ or $x_{n+1}=x_1$.

A link to integrable non-linear equations appears if we take
$$
u(\theta)={t\over 4}e^\theta+{\bar t\over 4}e^{-\theta}+r(\theta)
\eqno(3.9)
$$
with some $r(\theta)$ and denote
$$
U(u|\lambda)=W(u|\lambda)-W(u|-\lambda)=\log{D(u|-\lambda)\over D(u|\lambda)}
\eqno(3.10)
$$
Then it happens that (we denote $\partial=\partial/\partial t$ and $\bar
\partial=\partial/\partial\bar t$)
$$
4\partial\bar\partial W(u|\lambda)={1\over 4}\left(e^{2U(u|\lambda)}-1\right)
\eqno(3.11)
$$
or in terms of the determinants $D(u|\lambda)$
$$
D(u|\lambda)\partial\bar\partial D(u|\lambda)-\partial D(u|\lambda)\bar
\partial D(u|\lambda)={1\over 16}\left(D^2(u|\lambda)-D^2(u|-\lambda)\right)
\eqno(3.12)
$$
In particular $U=U(u|\lambda)=-U(u|-\lambda)$ solves the 2D sinh-Gordon
equation
$$
4\partial\bar\partial U={1\over 2}\sinh 2U
\eqno(3.13)
$$
Eq.(3.11) can be verified comparing order by order the $\lambda$-expansions in
the right and left hand sides. This would require the following relations
between the integrals (3.8)
$$
{16\over n}\partial\bar\partial W_n(u)=
\sum_{p_1,p_3,p_5,\ldots\ge 0\brace p_1+3p_3+5p_5+\ldots=n}\prod_{k=0}^\infty
{1\over p_{2k+1}!}\left({4W_{2k+1}(u)\over 2k+1}\right)^{p_{2k+1}}
\eqno(3.14)
$$
E.g.,
$$
\eqalign{
4\partial\bar\partial W_1&=W_1\cr
4\partial\bar\partial W_2&=4W_1^2\cr
4\partial\bar\partial W_3&=8W_1^3+W_3\cr
4\partial\bar\partial W_4&={32\over 3}W_1^4+{16\over 3}W_1W_3
}\eqno(3.15)
$$
etc. The first three relations in (3.15) are verified directly by comparing the
corresponding integrands. Starting from $n=4$ one has to perform a
symmetrisation in the rational in $x_i$, $i=1,2,\ldots,n$ part of the
integrands (this is allowed due to the permutation symmetry of the rest part
$\prod_{i=1}^n\exp(-2u(x_i))dx_i$, common for the both sides of eq.(3.14)).
I have checked (3.14) up to $n=6$. The complete proof perhaps can be found
following the lines of ref.[11]. It should be stressed that the validity of
(3.14) is of purely combinatorial nature, being based on certain algebraic
identities between the symmetrized rational expressions in $x_i$. Therefore we
expect eqs.(3.11--14) to hold independently on $\lambda$ and even on the
choice of the ``residual'' potential $r(\theta)$ in eq.(3.9). Moreover, they
remain valid if in eq.(3.1) (and therefore also in eq.(3.8)) we replace the
whole real axis by any other sensible integration contour in $\theta$. In this
note we shall not develop further these lines of generalizations.

Before turning at the rotationally symmetric (isotropic) version of eq.(3.11)
it is worth to mention the following interesting observation of refs.[1,2].
Consider a small variation $\delta u(\theta)$ of the potential in (3.2).
Then
$$
\delta W(u|\lambda)=-2\lambda\int R(\theta|\lambda)\delta u(\theta)d\theta
\eqno(3.16)
$$
where $R(\theta|\lambda)={\cal R}(\theta,\theta|\lambda)$ and ${\cal R}
(\theta,\theta'|\lambda)={\cal R}(\theta',\theta|\lambda)$ is the resolvent
kernel of eq.(3.1), i.e., the (unique at $\lambda\ne\lambda_a$) solution to
$$
{\cal R}(\theta,\theta'|\lambda)-\lambda\int K(\theta,\theta''){\cal R}
(\theta'',\theta'|\lambda)d\theta''=K(\theta,\theta')
\eqno(3.17)
$$
As a series in $\lambda$ it reads
$$
R(\theta|\lambda)=\sum_{n=0}^\infty\lambda^n R_n(\theta)
\eqno(3.18)
$$
where
$$
R_n(\theta)=e^{-2u(\theta)}\int{\displaystyle e^{-2u(\theta_1)-\ldots-
2u(\theta_n)}\over\displaystyle\cosh{\theta-\theta_1\over 2}\cosh{\theta_1-
\theta_2\over 2}\ldots\cosh{\theta_n-\theta\over 2}}d\theta_1\ldots d\theta_n
\eqno(3.19)
$$
We define also
$$
\eqalign{
2R_+(\theta|\lambda)&=R(\theta|\lambda)+R(\theta|-\lambda)=2\sum_{k=0}^\infty
\lambda^{2k}R_{2k}(\theta)\cr
2R_-(\theta|\lambda)&=R(\theta|\lambda)-R(\theta|-\lambda)=2\sum_{k=0}^\infty
\lambda^{2k+1}R_{2k+1}(\theta)
}\eqno(3.20)
$$
In refs.[1,2] on the basis of the field theory considerations quantity
$R_+(\theta|\lambda)$ was related to the following ``TBA-like'' system of
non-linear integral equations
$$
\eqalignno{
2u(\theta)&=\varepsilon(\theta)+\int_{-\infty}^\infty{\log(1+\eta^2(\theta'))
\over\cosh(\theta-\theta')}{d\theta'\over 2\pi}
&(3.21a)\cr
\eta(\theta)&=2\lambda\int{\displaystyle e^{-\varepsilon(\theta')}\over
\cosh(\theta-\theta')}d\theta'
&(3.21b)}
$$
with the same potential $u(\theta)$ and the same $\lambda$. It was argued that
there is a solution $\varepsilon(\theta)$, $\eta(\theta)$ to (3.21) such that
$$
R_+(\theta|\lambda)=e^{-\varepsilon(\theta)}
\eqno(3.22)
$$
In sect.6 of this note we extend slightly the TBA considerations of [1,2] to
argue a similar formula for $R_-(\theta|\lambda)$
$$
R_-(\theta|\lambda)=R_+(\theta|\lambda)\int_{-\infty}^\infty{\arctan
\eta(\theta')\over\cosh^2(\theta-\theta')}{d\theta'\over\pi}
\eqno(3.23)
$$

Strictly speaking, the arguments of [1,2] (and also of sect.6 below) are
relevant only for the isotropic case $2u(\theta)=t\cosh\theta$. However, if
we start to solve the system (3.21) as a power series in $\lambda$ we would
encounter for the corresponding terms in (3.22) and (3.23) expressions like
$$
\int e^{-2u(\theta_1)-\ldots-2u(\theta_n)}I(\theta_1,\ldots,\theta_n)
d\theta_1\ldots d\theta_n
\eqno(3.24)
$$
Here the ``potential weighted'' integrations over $\theta_1,\ldots,\theta_n$
arise from the iterations of eq.(3.21b) while $I(\theta_1,\ldots,\theta_n)$
results from the ``intermediate'' integrations in the first equation (3.21a)
or in eq.(3.23). They turn to be rational functions of $x_i=\exp\theta_i$;
$i=1,2,\ldots,n$ which, after symmetrization, are expected to reproduce the
(symmetrized) rational parts in the integrands of (3.19). In first few orders
this can be verified explicitly. If (3.21--23) is true in the isotropic case
(as it was argued in [1,2] and sect.6) this phenomenon must persist at all
orders, since it is quite hard to imagine that the transcendental integrations
over $\theta_1,\ldots,\theta_n$ may play any role in relating (3.24) to (3.19).
One concludes that the TBA representation (3.21--23) is of more general nature
and must hold for any reasonable potential $u(\theta)$. Also, corresponding to
different possible choices of the integration contour in (3.1) we have to
change the contour in (3.21b). Since the $\theta$-integrations in eqs.(3.21a)
and (3.23) are important in the building of $I(\theta_1,\ldots,\theta_n)$, the
contours there must remain unchanged. Of course, the consideration above is
not a proof, which is lacking up to now. I have verified the TBA representation
numerically for several (presumably random) samples of $u(\theta)$.

The TBA representation (although unproven) turns extremely useful in the
numerical calculations. E.g., for the isotropic case (3.25) with $|\lambda|<
2\pi$ (and not too close to $2\pi$) the iterative solution of the system
(3.21--23) converges fast and works much better then both the Painlev\'e
equation and the direct calculation of the Fredholm determinant. This typically
persists for other samples of $u(\theta)$ with $|\lambda^{-1}|$ well below the
$\sup_a|\lambda_a^{-1}|$. In particular, the curves of figs.5--7 are computed
in very this way.

{}From now on we switch to the rotationally symmetric (i.e., invariant under
$t\to\Lambda t$, $\bar t\to\Lambda^{-1}\bar t$) solutions to (3.11--13).
This corresponds to taking in (3.9) $r(\theta)={\rm const}$ (which can be
absorbed by the spectral parameter) and keeping $(-\infty,\infty)$ as the
integration region in (3.1). Thus we restrict to $t=\bar t$ real and
$$
2u(\theta)=t\cosh\theta
\eqno(3.25)
$$
In eqs.(3.11--13) we can substitute $2\partial=2\bar\partial=d/dt$,
$4\partial\bar\partial=d^2/dt^2+t^{-1}d/dt$ so that (3.13) is reduced to the
Painlev\'e III equation (2.13). The series (3.7) turns also the large $t$
expansion and we have
$$
W(t|\lambda)=2\lambda K_0(t)+8\lambda^2\left[-\log t\int_t^\infty K_0^2(r)rdr+
\int_t^\infty K_0^2(r)r\log rdr\right]+O\left(\lambda^3 e^{-3t}\right)
\eqno(3.26)
$$
Considering real solutions we have to take $\lambda$ real. It is possible to
show that at real $t>0$ $\sup_a\lambda_a^{-1}<2\pi$. Therefore $W(t|\lambda)$
is regular at $t>0$ for $\lambda\le (2\pi)^{-1}$. At $2\pi\lambda>1$ the
Fredholm determinant $D(t|\lambda)$ has infinite number of zeroes on the
positive $t$ axis, $t=0$ being their condensation point [11]. We shall consider
the case $2\pi|\lambda|\le 1$ where all the functions appearing in
eqs.(3.6--13) are regular at $t>0$. Denote
$$
2\pi\lambda=\sin{\pi\sigma\over 2}
\eqno(3.27)
$$
with real $-1\le\sigma\le 1$. It is relatively easy to analyse the $t\to 0$
behavior of the determinant (3.4) with the potential (3.25). One finds
$$
W(t|\lambda)={\sigma(\sigma+2)\over 4}\log{8\over t}+B(\sigma)+
O\left(t^{2\pm 2\sigma}\right)
\eqno(3.28)
$$
where ($\psi(x)=\Gamma'(x)/\Gamma(x)$)
$$
B(\sigma)={1\over 4}\int_0^\sigma(1+x)\left[\psi\left({1+x\over 2}\right)+
\psi\left({-1-x\over 2}\right)-2\right]dx
\eqno(3.29)
$$
In particular, for the combination (3.10) we have
$$
U(t|\lambda)=\sigma\log{8\over t}-{1\over 2}\log\kappa(\sigma)+
O\left(t^{2\pm 2\sigma}\right)
\eqno(3.30)
$$
with
$$
\kappa(\sigma)=\exp(2B(-\sigma)-2B(\sigma))={\displaystyle\Gamma^2\left(
{1\over 2}-{\sigma\over 2}\right)\over\displaystyle\Gamma^2\left({1\over 2}+
{\sigma\over 2}\right)}
\eqno(3.31)
$$
Using (3.30) and eqs.(3.11--13) one can systematically recover the further
$t\to 0$ corrections to (3.28). They appear as
$$
W(t|\lambda)={(\sigma+1)^2-1\over 4}\log{8\over t}-{t^2\over 16}+
\sum_{m,n=0}^\infty B_{m,n}(\sigma)\left({t\over 8}\right)^{2\alpha_{m,n}}
\eqno(3.32)
$$
where $\alpha_{m,n}=(m+n)+\sigma(m-n)$. In this double series
$$
B_{0,0}(\sigma)=B(\sigma)
\eqno(3.33)
$$
and the next coefficients $B_{m,n}(\sigma)$ are found order by order from the
following relation
$$
{1\over 4}\sum_{m,n=0}^\infty\alpha_{m,n}^2B_{m,n}(\sigma)x^{2\alpha_{m,n}}=
{\displaystyle x^{2\alpha_{0,1}}\over\kappa(\sigma)}\exp\left(2
\sum_{m,n=0\atop (m,n)\ne(0,0)}^\infty\left[B_{m,n}(\sigma)-B_{n,m}(-\sigma)
\right]x^{2\alpha_{m,n}}\right)
\eqno(3.34)
$$
Obviously $B_{m,0}(\sigma)=0$ for $m>0$. Few first $B_{m,n}$'s are
$$
\eqalign{
B_{0,1}(\sigma)&={4\over\kappa(\sigma)(1-\sigma)^2}\cr
B_{1,1}(\sigma)&=-{8\over(1+\sigma)^2}\cr
B_{0,2}(\sigma)&={8\over\kappa^2(\sigma)(1-\sigma)^4}
}\eqno(3.35)
$$

The polymer one-loop scaling functions of sect.2 correspond to the special case
$\sigma=1/3$.

\hbox{} \hbox{}
\vskip 0.5cm

{\bf 4. Relation to sin-Gordon}

The sin-Gordon model is a theory of 2D scalar field $\varphi(x)$,
$x^\mu=(x^0,x^1)$, with the action
$$
A_{sG}={1\over 2}\int(\partial_\mu\varphi)^2d^2x-\mu\int:\cos\beta\varphi:d^2x
\eqno(4.1)
$$
where $\beta$ is a real dimensionless parameter ($\beta^2\le 8\pi$). Here
$:\ldots:$ denotes the normal ordering with respect to massless free fields.
To be precise we choose such an infrared cutoff that at $\mu=0$ the two-point
correlation function $<\varphi(x)\varphi(0)>$ is
$$
\langle\varphi(x)\varphi(0)\rangle_{\mu=0}=-{1\over 2\pi}\log|x|
\eqno(4.2)
$$
In this scheme the (real) coupling constant $\mu$ is dimensional
$\mu\sim [{\rm mass}]^{2-2\Delta}$ with
$$
\Delta={\beta^2\over 8\pi}
\eqno(4.3)
$$
Sometimes it is convenient to use another (positive) parameter
$$
p={\beta^2\over 8\pi-\beta^2}
\eqno(4.4)
$$
instead of $\beta$.

The action (4.1) is obviously symmetric under the global shifts of the field
$\varphi$
$$
\varphi(x)\to\varphi(x)+{2\pi n\over\beta}
\eqno(4.5)
$$
where $n$ is an $x$-independent integer. Therefore the states $|\Psi>$ of the
model can be classified in their behavior under (4.5). The irreducible states
$|\Psi_\alpha>$ are characterized by the index (quasimomentum) $-\beta/2<
\alpha\le\beta/2$ such that
$$
\Psi_\alpha\left(\varphi+{2\pi\over\beta}\right)=e^{2i\pi\alpha/\beta}
\Psi_\alpha(\varphi)
\eqno(4.6)
$$
For each $\alpha$ there is an infinite dimensional space of states
${\cal A}_\alpha$, the complete space being
$$
{\cal A}=\bigoplus_{-\beta/2<\alpha\le\beta/2}{\cal A}_\alpha
\eqno(4.7)
$$
In the infinite volume system the transitions (4.5) are suppressed to zero so
that all the spaces ${\cal A}_\alpha$ are completely degenerate in $\alpha$.
However in a finite geometry (we imply that the finite geometry settlement
respects the symmetry (4.5)) these transitions are allowed and $\alpha$ becomes
an important parameter of the theory. Let us call ${\cal A}_\alpha$ the
$\alpha$-sector and the corresponding ground state $|\Psi_\alpha^{(0)}>$ the
$\alpha$-ground state. The $\alpha$-ground state energy $E_\alpha$ is a
periodic function of the quasimomentum $\alpha\sim\alpha+\beta\IZ$. For
symmetry reasons we expect $E_\alpha$ to have a minimum at $\alpha=0$ (mod
$\beta\IZ$), i.e. $|\Psi_0^{(0)}>$ is the true ground state.

Before considering the finite geometry effects remind briefly the well known
structure of the infinite volume space ${\cal A}_\alpha$ which is essentially
independent on $\alpha$. ${\cal A}_\alpha$ contains the infinite volume
ground state (the $\alpha$-vacuum) together with the excitations which are
(massive at $\mu\ne 0$ in (4.1)) relativistic particles subject a to factorized
scattering. The spectrum of particles always contains the soliton-antisoliton
doublet $(s,\bar s)$. Its mass $M$ is related to the coupling $\mu$ in eq.(4.1)
as follows [8]
$$
\pi\mu={\displaystyle\Gamma\left({p\over p+1}\right)\over\displaystyle
\Gamma\left({1\over p+1}\right)}\left[{\displaystyle\sqrt{\pi}M\Gamma\left(
{p+1\over 2}\right)\over\displaystyle 2\Gamma\left({p\over 2}\right)}
\right]^{2/(p+1)}
\eqno(4.8)
$$
Quasiclassically one can think about $s$ and $\bar s$ as of the kink
configurations of field $\varphi(x)$ such that (currently we imply $x^1$ to
be the spatial coordinate)
$$
\varphi(x^1\to\infty)-\varphi(x^1\to -\infty)=\pm 2\pi/\beta
\eqno(4.9)
$$ for the soliton and antisoliton respectively. At $p<1$ the spectrum includes
also the $s-\bar s$ bound states $B_n$, $n=1,2,\ldots<1/p$ of masses
$$
m_n=2M\sin{\pi pn\over 2}
\eqno(4.10)
$$
The factorized scattering amplitudes of all these particles can be found e.g.
in [12]. Note that all the observables are independent on $\alpha$. The
(specific) ground state energy [7,8]
$$
{\cal E}=-{M^2\over 4}\tan{\pi p\over 2}
\eqno(4.11)
$$
is also $\alpha$ independent.

Let us now put the sin-Gordon model on a finite space circle of circumference
$R$ and impose the periodic boundary conditions, i.e., $\varphi(x^0,x^1+R)=
\varphi(x^0,x^1)$. In the euclidean version of (4.1) this corresponds to the
geometry of infinite (or very long $L\to\infty$ in the ``time'' direction
$x^0$) flat cylinder. We denote $E_\alpha(R)$ the corresponding $\alpha$-ground
state energy. In the perturbed conformal field theory (CFT) picture this
observable would correspond to the specific (per unit length of the cylinder)
free energy with the scalar operator $\exp(i\alpha\varphi)$ ``flowing
along the cylinder''. The UV conformal dimension of this operator is
$\Delta_\alpha=\alpha^2/8\pi$. Therefore we expect at $R\to 0$ [13]
$$
E_\alpha(R)\sim -{\pi c_\alpha\over 6R}=-{\pi\over 6R}\left(1-{3\alpha^2\over
\pi}\right)
\eqno(4.12)
$$
In general
$$
E_\alpha(R)=MF_{sG}(\alpha,p|MR)
\eqno(4.13)
$$
where the soliton mass $M$ is used as an overall scale to isolate the scale
independent function $F_{sG}$ and the dependence on the sin-Gordon parameter
(4.4) is explicitly indicated. From (4.12) we have at $t\to 0$
$$
F_{sG}(\alpha,p|t)\sim -{\pi\over 6t}\left(1-{3\alpha^2\over\pi}\right),\ \ \
\ \ t\to 0
\eqno(4.14)
$$
independently on $p$. In the opposite limit $t\to\infty$ we expect from (4.11)
$$
F_{sG}(\alpha,p|t)=-{t\over 4}\tan{\pi p\over 2}+{\rm corrections},\ \ \ \ \ \
t\to\infty
\eqno(4.15)
$$
independently on $\alpha$. The leading correction in (4.15) comes (at $p>1/3$)
from the virtual soliton or antisoliton trajectory winding once around the
cylinder. In view of (4.9) it is plain that in the $\alpha$-sector these
trajectories are weighted by the factors of $\exp(\pm 2i\pi\alpha/\beta)$
respectively. Therefore, summing up the $s$ and $\bar s$ contributions
$$
F_{sG}(\alpha,p|t)+{t\over 4}\tan{\pi p\over 2}=-{2\over\pi}\cos{2\pi\alpha
\over\beta}K_1(t)+\hbox{(multiparticle or bound state contributions)}
\eqno(4.16)
$$

Considering at the microscopic level one can argue [14] that the described
above cylinder sin-Gordon settlement is in the same universality class as the
cylinder polymer counting problem of sect.2 provided the non-winding polymer
weight $n$ is related to the sin-Gordon parameter (4.4) by eq.(1.5). Also,
comparing (4.16) with what one expects at large $t$ for the polymer counting
scaling function $F(m,n|t)$ we can relate the sin-Gordon quasimomentum $2\pi
\alpha/\beta$ to the winding polymer weight $m$
$$
m=2\cos{2\pi\alpha\over\beta}
\eqno(4.17)
$$
In addition it is clear that the soliton mass $M$ has to be identified with the
inverse correlation length of the polymer problem. Altogether it seems not
misleading to think of the polymer loops as of the virtual soliton trajectories
(summed over the ``orientations'' $s$ and $\bar s$). Note, that the
sin-Gordon--polymer relation (1.5) implies that $p>1$ where the sin-Gordon
spectrum contains no extra particles but $s$ and $\bar s$. It looks quite
natural that the self-avoiding polymers form no bound states.

We conclude that if $m$ and $n$ are related to $p$ and $\alpha$ as in eqs.(1.5)
and (4.17)
$$
F(m,n|t)=F_{sG}(\alpha,p|t)
\eqno(4.18)
$$
This correspondence turns rather useful. It permits often to reinterpret the
sin-Gordon results in the polymer terms. E.g., eq.(1.7) is simply read off from
the ground state energy (4.11). Vise versa (4.18) implies that at $m=n=0$, i.e.
$p=2$ and $2\alpha/\beta=1/2$ the sin-Gordon $\alpha$-ground state energy
$F_{sG}(\alpha,p|t)$ vanishes for all $t$. This is indeed the case due to the
$N=2$ supersymmetry of the sin-Gordon model at this point [15]. One then can
expand $F_{sG}(\alpha,p|t)$ around this point to obtain the scaling functions
$F_{l_1,l_2}(t)$ in eq.(2.9).

The standard tools of studying the sin-Gordon model become applicable to the
polymer problem of sect.2. In particular, in the next section we use the
perturbation theory of (4.1) in $\mu$ to evaluate the small $t$ behavior of
$F(m,n|t)$. The perturbative series for $E_\alpha(R)$ is constructed in the
usual way. In fact it is the expansion
$$
E_\alpha(R)=-{\pi\over 6R}\sum_{n=0}^\infty c_{2n}(\alpha)\eta^{2n}
\eqno(4.19)
$$
in the dimensionless variable
$$
\eta=-2\pi\mu\left({R\over 2\pi}\right)^{2/(p+1)}=-2\pi r_p\left({t\over
2\pi}\right)^{2/(p+1)}
\eqno(4.20)
$$
where
$$
r_p=\mu M^{-2/(p+1)}
\eqno(4.21)
$$
is the ``$\mu$-$M$ ratio'' defined by eq.(4.8). The first coefficient in (4.19)
$$
c_0(\alpha)=1-{3\alpha^2\over\pi}
\eqno(4.22)
$$
is the effective central charge corresponding to operator
$\exp(i\alpha\varphi)$ while the subsequent ones are given by perturbative
integrals. Introducing the complex coordinate $z=x^1+ix^0$ on the cylinder
$z\sim z+R\IZ$ and substituting $u=\exp(2\pi iz/R)$ we have explicitly
$$
\eqalign{
&c_{2n}(\alpha)={12\over(2\pi)^{2n-1}(n!)^2}\times\cr
&\int\limits_{(u_1=1)}\left[{\displaystyle
\prod_{i>j}^n|u_i-u_j|^{4\Delta}|v_i-v_j|^{4\Delta}\over\displaystyle
\prod_{i,j}^n|u_i-v_j|^{4\Delta}}\prod_{i=1}^n|u_i|^{2\Delta(1-2\alpha/\beta)}
|v_i|^{2\Delta(1+2\alpha/\beta)}-{\hbox{subtrac-}\atop\hbox{tions}}\right]
{d^2u_i\over|u_i|^2}{d^2v_i\over|v_i|^2}
}\eqno(4.23)
$$
where the disconnected parts are subtracted to build the connected $2n$-point
cylinder correlation function.
The integrand is invariant under the homogeneous shifts along the cylinder
$u_i\to\Lambda u_i$, $v_i\to\Lambda v_i$ so that we can fix one of the points
($u_1$ in eq.(4.23)). Note that the perturbative coefficients $c_{2n}(\alpha)$
are not periodic in the quasimomentum $\alpha$. In fact expansion (4.19)
determines the $\alpha$-ground state energy only if $|\alpha|<\beta/2$. For
$|\alpha|>\beta/2$ and $\alpha=\alpha_0$ (mod $\beta$), $|\alpha_0|<\beta/2$ it
corresponds instead to an excited state in the $\alpha_0$ sector. The
discontinuities occur at $\alpha=n\beta/2$ with $n$ integer, where the
unperturbed space of states contains two (connected perturbatively) degenerate
states $\exp(\pm in\beta\varphi/2)$ and expansion (4.19) has to be modified.

The series (4.19) admits another interpretation if $2\alpha/\beta=\pm(1/p-k)$
with $k$ a non-negative integer [16]
(note that in this series only $k=0$ and $1$
would correspond to $\alpha$-ground states). Take e.g., $2\alpha/\beta=1/p-k$
and consider first the integration over $v_i$, $i=1,2,\ldots,n$ in eq.(4.23).
Following [17] we have
$$
\eqalign{
&{1\over n!}\int{\displaystyle\prod_{i>j}^n|u_i-u_j|^{4p/(p+1)}|v_i-v_j|^{4p/
(p+1)}\over\displaystyle\prod_{i,j}^n|u_i-v_j|^{4p/(p+1)}}\prod_{i=1}^n
|u_i|^{2(kp-2)/(p+1)}|v_i|^{-2kp/(p+1)}d^2v_i=\cr
&\left({2\pi\over Q_p}\right)^n\langle\Phi_{1,k+1}(0)\Phi_{1,3}(u_1)\Phi_{1,3}
(u_2)\ldots\Phi_{1,3}(u_n)\Phi_{1,k+1}(\infty)\rangle_{{\cal M}_p}
\prod_{i=1}^n|u_i|^{-4/(p+1)}
}\eqno(4.24)
$$
where $\Phi_{1,k}$, $k=1,2,\ldots$ denote the (thermal series) primary fields
in the minimal CFT ${\cal M}_p$ and $\langle\ldots\rangle_{{\cal M}_p}$ is
quite the correlation function in this model. Corresponding to the usual in CFT
normalization of fields $<\Phi_{1,k}(x)\Phi_{1,k}(0)>_{{\cal M}_p}=|x|^{-4
\Delta_k}$ (here $4\Delta_k=k(kp-2)/(p+1)$) we have to take
$$
Q_p={2(2p-1)(p-1)\over (p+1)^2}\left[{\displaystyle\Gamma^3\left({p\over p+1}
\right)\Gamma\left({1-2p\over p+1}\right)\over\displaystyle\Gamma^3\left(
{1\over p+1}\right)\Gamma\left({3p\over p+1}\right)}\right]^{1/2}
\eqno(4.25)
$$
Thus denoting
$$
\lambda=-{2\pi\mu^2\over Q_p}
\eqno(4.26)
$$
one can interpret the series (4.19) as the perturbative expansion in the
$\Phi_{13}$-perturbed ${\cal M}_p$, i.e., in the RSG model
$$
A_\lambda=A_{{\cal M}_p}+\lambda\int\Phi_{13}(x)\, d^2x
\eqno(4.27)
$$
where $A_{{\cal M}_p}$ is the formal action of the CFT ${\cal M}_p$. Under
this interpretation the quantity (4.19) at $2\alpha/\beta=1/p-k$ becomes the
perturbed finite size energy of the primary state $\Phi_{1,k+1}$ in the model
(4.27). In particular $k=0$ and $k=1$ correspond to the perturbed states
$\Phi_{11}=\II$ and $\Phi_{12}$ respectively. If $2\alpha/\beta=-1/p+k$
we can similarly consider first the integration over $u_i$ in (4.23) and then
treat the remaining integral over $v_i$ as a perturbative one in the model
(4.27).

Let us denote $F_{(k)}(p|t)$ the scaling function corresponding to the finite
size energy of the RSG state $\Phi_{1,k+1}$. If the sin-Gordon coupling
$\mu$ is related to $\lambda$ as in eq.(4.26) we have
$$
F_{(k)}(p|t)=F_{sG}(\pm(1/p-k),p|t)
\eqno(4.28)
$$
Comparing with (4.18) we see that the polymer functions (2.7) and (2.8) are
(provided relation (1.5) between $p$ and $n$ holds)
$$
\eqalign{
F_+(n|t)&=F_{(0)}(p|t)\cr
F_-(n|t)&=F_{(1)}(p|t)
}\eqno(4.29)
$$
i.e. correspond to the RSG states $\II$ and $\Phi_{12}$ respectively.

\hbox{}
\vskip 0.5cm

{\bf 5. Perturbation theory}

Here we present few explicit calculations about the sin-Gordon perturbative
series (4.19). For the later comparisons it is convenient to use the
scaling parameter
$$
\xi=(2\pi)^{-2(p-1)/(p+1)}\eta^2/r_p^2
\eqno(5.1)
$$
instead of $\eta$ in eq.(4.19) (here $r_p$ is the $\mu$-$M$ ratio (4.8),
(4.12)). At $p=2$ this notation conforms that of eq.(2.17). From (4.19) we
have
$$
F_{sG}(\alpha,p|t)=-{\pi\over 6t}\sum_{n=0}^\infty c_{2n}(\alpha)\left[
(2\pi)^{(p-1)/(p+1)}r_p\right]^{2n}\xi^n
\eqno(5.2)
$$
At $n=1$ the integral (4.23) is carried out explicitly
$$
c_2(\alpha)=6{\displaystyle\gamma\left({p\over p+1}\left(1-{2\alpha\over
\beta}\right)\right)\gamma\left({p\over p+1}\left(1+{2\alpha\over\beta}
\right)\right)\over\displaystyle\gamma\left({2p\over p+1}\right)}
\eqno(5.3)
$$
where the notation $\gamma(x)=\Gamma(x)/\Gamma(1-x)$ is used.

In the polymer language eq.(5.2) together with (4.22) and (5.3) predicts the
leading $t\to 0$ behavior and the next-to-leading correction to the polymer
scaling function $F(m,n|t)$. In particular, expanding around $p=2$ and
$2\alpha/\beta=1/2$ we find
$$
tF(m,n|t)=m\left(-{1\over 3}+{3\over 4\gamma^2(1/3)}\xi+\ldots\right)+
n\left({1\over 18}-{3\over 4\gamma^2(1/3)}\xi+\ldots\right)+\ldots
\eqno(5.4)
$$
in agreement with eqs.(2.19) and (2.30).

At $2\alpha/\beta=1/p$ (this value
corresponds to $m=n$ in the polymer language) we can go two steps further
thanks to the RSG interpretation described in sect.4. Namely
$$
\eqalign{
c_4\left({2\alpha\over\beta}={1\over p}\right)&={3\over\pi Q^2_p}
\int\langle\Phi_{13}(1)\Phi_{13}(u)\rangle_{{\cal M}_p}|u|^{-4/(p+1)}d^2u\cr
&={\displaystyle 3(p+1)^4\gamma^2\left({p-1\over p+1}\right)\gamma\left(
{3p\over p+1}\right)\over\displaystyle 4(p-1)^2(2p-1)^2\gamma\left(
{2p-2\over p+1}\right)\gamma^3\left({p\over p+1}\right)}
}\eqno(5.5)
$$
and
$$
\eqalign{
c_6\left({2\alpha\over\beta}={1\over p}\right)&={1\over 2\pi^2 Q^3_p}\int
\langle\Phi_{13}(1)\Phi_{13}(u_1)\Phi_{13}(u_2)\rangle_{{\cal M}_p}
|u_1u_2|^{-4/(p+1)}d^2u_1d^2u_2\cr
&=-{\displaystyle 18(p+1)^4\gamma\left({p-1\over p+1}\right)\gamma^2\left(
{3p\over p+1}\right)\gamma\left({2p-2\over p+1}\right)\over\displaystyle
(p-1)^2(2p-1)^2\gamma\left({5p-1\over p+1}\right)\gamma^6\left({p\over p+1}
\right)}
}\eqno(5.6)
$$
At $p=2+2n/\pi+O(n^2)$ this gives for the RSG scaling function
$F_{(0)}(p|t)$ (see eq.(4.28))
$$
tF_{(0)}(p|t)=n\left(-{5\over 18}-{3\gamma^2(1/3)\over 64}\xi^2+
{9\over 256}\xi^3+\ldots\right)+O(n^2)
\eqno(5.7)
$$
in complete accordance with the expansion (2.31).

Another special case of the oppositely weighted winding and non-winding
polymer loops $m=-n$ is related by eq.(4.17) to the sin-Gordon quasimomentum
$2\alpha/\beta=\pm(1/p-1)$. This point again admits the RSG
interpretation corresponding now to the finite-size ground state in the
$\Phi_{12}$ sector of the $\Phi_{13}$-perturbed ${\cal M}_p$. Therefore
$$
c_4\left({2\alpha\over\beta}={1\over p}-1\right)={3\over\pi Q_p^2}\int
\ll\Phi_{12}(0)\Phi_{13}(1)\Phi_{13}(u)\Phi_{12}(\infty)
\gg_{{\cal M}_p}{d^2u\over|u|^2}
\eqno(5.8)
$$
where $\ll\ldots\gg_{{\cal M}_p}$ denotes the connected cylinder correlation
function in ${\cal M}_p$. Explicitly
$$
\ll\Phi_{12}(0)\Phi_{13}(1)\Phi_{13}(u)\Phi_{12}(\infty)
\gg_{{\cal M}_p}=a_1\left(f_1(u)f_1(\bar u)-1\right)+a_2f_2(u)f_2(\bar u)
\eqno(5.9)
$$
where
$$
a_1={\displaystyle\gamma\left({p\over p+1}\right)\gamma\left({2p-1\over p+1}
\right)\over\displaystyle\gamma\left({2p\over p+1}\right)\gamma\left(
{p-1\over p+1}\right)}\ ;\ \ \ \
a_2={\displaystyle\gamma\left({p\over p+1}\right)\gamma\left({1-2p\over p+1}
\right)\over\displaystyle\gamma\left({2-2p\over p+1}\right)\gamma\left(
{p-1\over p+1}\right)}
\eqno(5.10)
$$
and $f_{1,2}$ are in terms of the hypergeometric functions
$$
\eqalign{
f_1(u)&=\displaystyle (1-u)^{(2-2p)/(p+1)}F\left({2-2p\over p+1},{1\over p+1},
{2-p\over p+1},u\right)\cr
f_2(u)&=\displaystyle u^{(2p-1)/(p+1)}(1-u)^{(2-2p)/(p+1)}
F\left({2p\over p+1},{1\over p+1},{3p\over p+1},u\right)
}\eqno(5.11)
$$
The integral (5.8) again can be carried out
$$
\eqalign{
&c_4\left({2\alpha\over\beta}={1\over p}-1\right)=
{\displaystyle
3\gamma^2\left({2p-1\over p+1}\right)\over\displaystyle 2\gamma^2\left(
{2p\over p+1}\right)\gamma^2\left({p\over p+1}\right)}
\left[\psi\left({2-p\over p+1}\right)\right.+\cr
&\left.\psi\left({2p-1\over p+1}\right)
-\psi\left({p\over p+1}\right)-\psi\left({1\over p+1}\right)-\psi\left(
{3p-1\over 2(p+1)}\right)-\psi\left({3-p\over 2(p+1)}\right)-2\psi\left(
{1\over 2}\right)\right]
}\eqno(5.12)
$$
At $p\to 2$ we have therefore
$$
tF_{(1)}(p|t)=n\left({7\over 18}-{3\over 2\gamma^2(1/3)}\xi-{27\over
32\gamma^4(1/3)}\xi^2+\ldots\right)+O(n^2)
\eqno(5.13)
$$
to be compared with the first terms of eq.(2.32).

\hbox{}
\vskip 0.5cm

{\bf 6. TBA considerations}

In the previous two sections the sin-Gordon scaling function $F_{sG}(\alpha,
p|t)$ has been studied in the UV perturbation theory. In principle this
approach provides us with a systematic expansion of $F_{sG}(\alpha,p|t)$ in
powers of $t^{2/(p+1)}$. At the same time the integrability of the
sin-Gordon model allows the same observable $F_{sG}$ to be considered in the
TBA framework. In TBA one does not concern directly the field theory action
starting instead from the (exactly known in the sin-Gordon case) relativistic
factorized scattering theory. After some manipulations the cylinder scaling
function is related to a system of non-linear integral equations (the TBA
system), its form depending on the factorized scattering amplitudes.

In the case of the sin-Gordon scattering theory the TBA system (at $p>1$) is
borrowed essentially from the construction by Takahashi and Suzuki [18] for
the anisotropic Heisenberg chain. The structure depends drastically on the
arithmetic nature of the sin-Gordon parameter $p$. In general it is an
infinite system of coupled non-linear integral equations for infinite number
of unknown functions. At $p$ rational it can be simplified and reduced to a
finite system (see [18] for the details). This is why in the TBA studies of
the sin-Gordon model one tends to choose $p$ rational. It should be mentioned
however that this ``fractal'' dependence on $p$ is a problem of the TBA
technique itself. The sin-Gordon physics is quite continuous in the coupling
constant.

For the time being we are interested in the point $p=2$ and its vicinity. At
$p=2$ the TBA system contains only two unknown functions $e_1(\theta)$ and
$e_2(\theta)$ and reads
$$
\eqalign{
t\cosh\theta&=e_1+s*\log\left(1+2\cos\left(4\pi\alpha/\beta\right)
e^{-e_2}+e^{-2e_2}\right)\cr
0&=e_2+s*\log\left(1+e^{-e_1}\right)
}\eqno(6.1)
$$
where $*$ denotes the convolution in $\theta$
$$
s*g=\int_{-\infty}^{\infty}s(\theta-\theta')g(\theta')d\theta'
\eqno(6.2)
$$
with the kernel
$$
s(\theta)=\int{\displaystyle e^{i\omega\theta}\over\displaystyle 2
\cosh{\pi\omega\over 2}}{d\omega\over 2\pi}={1\over 2\pi\cosh\theta}
\eqno(6.3)
$$
The sin-Gordon scaling function appears as
$$
F_{sG}(\alpha,2|t)=-{1\over 2\pi}\int\cosh\theta\log\left(1+e^{-e_1(\theta)}
\right)d\theta
\eqno(6.4)
$$
It happens that at $m=2\cos(2\pi\alpha/\beta)$ small, $e_1\sim -\log m$ and
$e_2\sim m$. Denoting
$$
\eqalign{
e^{-e_1(\theta)}&=me^{-\varepsilon(\theta)}\cr
e^{-e_2(\theta)}&=1+m\eta(\theta)
}\eqno(6.5)
$$
we find that at $m\to 0$
$$
F_{sG}(\alpha,2|t)=-m\int\cosh\theta e^{-\varepsilon(\theta)}{d\theta\over
2\pi}+O(m^2)
\eqno(6.6)
$$
while $\varepsilon(\theta)$ and $\eta(\theta)$ solve the following system
$$
\eqalign{
t\cosh\theta&=\varepsilon+s*\log(1+\eta^2)\cr
\eta&=s*e^{-\varepsilon}
}\eqno(6.7)
$$
which is quite the one quoted in sect.3 eq.(3.21) (with $\lambda=1/4\pi$).
Note, that eqs.(6.7) imply the following functional system for $R_+(\theta)=
\exp (-\varepsilon(\theta))$ and $\eta(\theta)$
$$
\eqalign{
R_+(\theta+i\pi/2)R_+(\theta-i\pi/2)&=1+\eta^2(\theta)\cr
\eta(\theta+i\pi/2)+\eta(\theta-i\pi/2)&=R_+(\theta)
}\eqno(6.8)
$$
Moreover, it can be shown that under appropriate analytic and asymptotic
restrictions on $R_+(\theta)$ and $\eta(\theta)$ the functional and
integral systems (6.8) and (6.7) are equivalent.

To arrive at eq.(3.23) one needs to shift slightly from the point $p=2$. For
the reasons mentioned above and following ref.[1] we choose the series of
rational values
$$
p=2+1/N\ ,\ \ \ \ N=2,3,4,\ldots
\eqno(6.9)
$$
Any analytic at $p=2$ function of $p$ is unambiguously recovered from the
series of its values at (6.9).

The Takahashi-Suzuki TBA system at $p=2+1/N$ is quoted in [1]. It includes
$N+2$ functions $\varepsilon_a(\theta)$; $a=0,1,\ldots,N+1$ and has the form
$$
\eqalign{
t\cosh\theta&=\varepsilon_0+s*L_1\cr
0&=\varepsilon_1+s*L_0+s_N*L_2+d_N*L_1\cr
0&=\varepsilon_2+s_N*[L_3-L_1]\cr
3\le a\le N,\ \ \ 0&=\varepsilon_a+s_N*[L_{a+1}+L_{a-1}]\cr
0&=\varepsilon_{N+1}+s_N*L_N
}\eqno(6.10)
$$
The kernel $s(\theta)$ is given by eq.(6.3),
$$
\eqalign{
s_N(\theta)&=\int{\displaystyle e^{i\omega\theta}\over\displaystyle
2\cosh{\pi\omega\over 2N}}{d\omega\over 2\pi}\cr
d_N(\theta)&=\int{\displaystyle e^{i\omega\theta}\cosh{\pi(N-1)\omega\over 2N}
\over\displaystyle 2\cosh{\pi\omega\over 2}\cosh{\pi\omega\over 2N}}
{d\omega\over 2\pi}
}\eqno(6.11)
$$
while the functions $L_a(\theta)$; $a=0,1,2,\ldots,N+1$ in eq.(6.10) are read
as follows
$$
\eqalign{
L_a(\theta)&=\log\left(1+e^{-\varepsilon_a(\theta)}\right),\ \ \ \ \ \ \
0\le a\le N\cr
L_{N+1}(\theta)&=\log\left[\left(1+qe^{-\varepsilon_{N+1}(\theta)}\right)
\left(1+q^{-1}e^{-\varepsilon_{N+1}(\theta)}\right)\right]
}\eqno(6.12)
$$
Here
$$
q=\exp\left({2i\pi\alpha\over\beta}(2N+1)\right)
\eqno(6.13)
$$
introduces the sin-Gordon quasimomentum $\alpha$. The cylinder scaling function
is determined by $L_0(\theta)$
$$
F_{sG}\left(\alpha,2+1/N|t\right)=-{1\over 2\pi}\int\cosh\theta
L_0(\theta)d\theta-{t\over 4}\tan{\pi\over 2N}
\eqno(6.14)
$$
For our purposes it is more convenient to consider the functional system
for $Y_a(\theta)=\exp(-\varepsilon_a(\theta))$, $a=0,1,2,\ldots,N+1$, which
follows from (6.10)
$$
Y_0\left(\theta+{i\pi\over 2}\right)Y_0\left(\theta-{i\pi\over 2}\right)=
1+Y_1(\theta)\ \ ;\phantom{XXX}
$$
$$
\eqalign{
&{\displaystyle Y_1\left(\theta+{i\pi(N+1)\over 2N}\right)Y_1\left(\theta-
{i\pi(N+1)\over 2N}\right)\over\displaystyle\left(1+Y_1^{-1}\left(
\theta+{i\pi(N-1)\over 2N}\right)\right)\left(1+Y_1^{-1}\left(\theta-
{i\pi(N-1)\over 2N}\right)\right)}=\cr
&\left(1+Y_0\left(\theta+{i\pi\over 2N}
\right)\right)\left(1+Y_0\left(\theta-{i\pi\over 2N}\right)\right)
\left(1+Y_2\left(\theta+{i\pi\over 2}\right)\right)\left(1+Y_2\left(\theta-
{i\pi\over 2}\right)\right)\ \ ;
}
$$
$$
\eqalign{
Y_2\left(\theta+{i\pi\over 2N}\right)Y_2\left(\theta-{i\pi\over 2N}\right)&=
{1+Y_3(\theta)\over 1+Y_1(\theta)}\ \ ;\cr
Y_a\left(\theta+{i\pi\over 2N}\right)Y_a\left(\theta-{i\pi\over 2N}\right)&=
\left(1+Y_{a+1}(\theta)\right)\left(1+Y_{a-1}(\theta)\right)\ \ ,\ \ \ \
2\le a\le N-1\ \ ;\cr
Y_N\left(\theta+{i\pi\over 2N}\right)Y_N\left(\theta-{i\pi\over 2N}\right)&=
\left(1+Y_{N-1}(\theta)\right)\left(1+qY_{N+1}(\theta)\right)\left(1+q^{-1}
Y_{N+1}(\theta)\right)\ \ ;\cr
Y_{N+1}\left(\theta+{i\pi\over 2N}\right)Y_{N+1}\left(\theta-{i\pi\over 2N}
\right)&=1+Y_N(\theta)\cr
}\eqno(6.15)
$$
It is easy to verify that this system is ``solved'' in terms of a single
function $G(\theta)$ (defined up to an overall multiplying constant) as
$$
\eqalign{
Y_0(\theta)&={\left(q^2G_{2N+1}(\theta)-q^{-2}G_{-2N-1}(\theta)\right)
\left(G_1(\theta)-G_{-1}(\theta)\right)\over\left(qG_{2N+1}(\theta)-
%% FOLLOWING LINE CANNOT BE BROKEN BEFORE 80 CHAR
q^{-1}G_1(\theta)\right)\left(qG_{-1}(\theta)-q^{-1}G_{-2N-1}(\theta)\right)}\cr
1+Y_1(\theta)&={\left(G_{N+1}(\theta)-G_{N-1}(\theta)\right)
\left(G_{-N+1}(\theta)-G_{-N-1}(\theta)\right)\over\left(qG_{N-1}(\theta)-
q^{-1}G_{-N-1}(\theta)\right)\left(qG_{N+1}(\theta)-q^{-1}G_{-N+1}(\theta)
\right)}\cr
Y_{N+1}(\theta)&=-{qG_1(\theta)-q^{-1}G_{-1}(\theta)\over G_1(\theta)-
G_{-1}(\theta)}\cr
}\eqno(6.16)
$$
and for $2\le a\le N$
$$
1+Y_a(\theta)={\left(qG_{N-a}(\theta)-q^{-1}G_{-N+a-2}(\theta)\right)
\left(qG_{N-a+2}(\theta)-q^{-1}G_{-N+a}(\theta)\right)\over
\left(G_{N-a+2}(\theta)-G_{N-a}(\theta)\right)\left(G_{-N+a}(\theta)-
G_{-N+a-2}(\theta)\right)}
\eqno(6.17)
$$
provided $G(\theta)$ satisfies
$$
G_{6N+2}(\theta)=q^{-6}G(\theta)
\eqno(6.18)
$$
In eqs.(6.16--18) we use the abbreviation
$$
G_k(\theta)=G\left(\theta+{i\pi k\over 2N}\right)
\eqno(6.19)
$$
Note that due to (6.18) all the functions $Y_a(\theta)$ are
$i\pi(3+1/N)$-periodic in $\theta$.

Turn at the limit $N\to\infty$ and $2\alpha/\beta=1/2-\epsilon/2$ with
$\epsilon\to 0$. Following eqs.(1.5) and (4.17)
$$
\eqalign{
m&=\pi\epsilon+O(\epsilon^2)\cr
n&={\pi\over 2N}+O\left({1\over N^2}\right)
}\eqno(6.20)
$$
In this notation $q=i(-)^N\exp(-i\pi\epsilon(N+1/2))$. It is convenient to
introduce new function
$$
h(\theta)=G(\theta)\exp\left(-{3N(2N+1)\epsilon\over 2N+1}\theta\right)
\eqno(6.21)
$$
which enjoys the $m,n$-independent behavior under the period shift (6.18)
$$
h(\theta+i\pi(3+1/N))=-h(\theta)
\eqno(6.22)
$$
The function $Y_0(\theta)$, which is most important for us, now becomes (again
$h_k(\theta)=h(\theta+i\pi k/2N)$)
$$
Y_0(\theta)={\left(e^{i\delta/3}h_{2N+1}(\theta)-e^{-i\delta/3}
h_{-2N-1}(\theta)\right)\left(e^{i\delta}h_1(\theta)-e^{-i\delta}
h_{-1}(\theta)\right)\over\left(e^{i\delta/3}h_{2N+1}(\theta)+
e^{i\delta}h_1(\theta)\right)\left(e^{-i\delta}h_{-1}(\theta)+
e^{-i\delta/3}h_{-2N-1}(\theta)\right)}
\eqno(6.23)
$$
with $\delta=\pi\epsilon(N+1/2)/(N+1/3)$. At $N=\infty$ and $\epsilon=0$ this
quantity vanishes as one could expect, while $h(\theta)$ turns to some
limiting function $g(\theta)$ such that
$$
g(\theta+3i\pi)=-g(\theta)
\eqno(6.24)
$$
Moreover, the first order terms in $m$ and $n$
$$
Y_0(\theta)=mR_+(\theta)-nR_-(\theta)+\hbox{higher order terms}
\eqno(6.25)
$$
are determined by $g(\theta)$
$$
\eqalignno{
R_+(\theta)&={2i\left(g(\theta+i\pi)-g(\theta-i\pi)\right)g(\theta)\over
\left(g(\theta+i\pi)+g(\theta)\right)\left(g(\theta)+g(\theta-i\pi)\right)}
&(6.26a)\cr
R_-(\theta)&=-{2i\left(g(\theta+i\pi)-g(\theta-i\pi)\right)g'(\theta)\over
\left(g(\theta+i\pi)+g(\theta)\right)\left(g(\theta)+g(\theta-i\pi)\right)}
&(6.26b)\cr
}
$$
Denoting also
$$
\eta(\theta)=-i{g(\theta+i\pi/2)-g(\theta-i\pi/2)\over g(\theta+i\pi/2)+
g(\theta-i\pi/2)}
\eqno(6.27)
$$
one finds that (6.26a) and (6.27) satisfy the functional system (6.8).
Therefore we have to identify these functions with the solutions to the TBA
system (6.7).

{}From (6.27) one can find $g(\theta)$ in terms of $\eta(\theta)$
$$
\log g(\theta)=-{1\over\pi}\int\tanh(\theta-\theta')\arctan\eta(\theta')
d\theta'+\hbox{const}
\eqno(6.28)
$$
Relation (3.23) follows.

The TBA system (6.7) together with (3.23) has been used to compute numerically
$$
F_n(t)=-{1\over 2\pi}\int\cosh\theta R_+(\theta)d\theta
\eqno(6.29)
$$
and
$$
F_c(t)+{t\over 4}={1\over 2\pi}\int\cosh\theta R_-(\theta)d\theta
\eqno(6.30)
$$
for the plots of figs.5--7.

\hbox{}
\vskip 0.5cm

{\bf 7. Concluding remarks}

Expansions (2.19) and (2.28) imply that the scaling functions $tF_n(t)$ and
$tF_c(t)$ are still real (at least at $|t|$ small enough) after the analytic
continuation
$$
t\to e^{3i\pi/4}t
\eqno(7.1)
$$
(or $\xi\to -\xi$ in eq.(2.1)). As it is argued in [9] such an analytic
continuation would correspond to the change from the subcritical (dilute)
polymer phase to the supercritical (dense) scaling polymers. The continued
functions $\tilde F_n(t)=e^{3i\pi/4}F_n(e^{3i\pi/4}t)$ and $\tilde F_c(t)=
e^{3i\pi/4}F_c(e^{3i\pi/4}t)$ are given (inside the convergence region) by the
alternated series (2.19) and (2.28) (again $\xi=t^{4/3}$)
$$
\eqalign{
t\tilde F_n(t)&=-{1\over 3}+{4\over 3}\sum_{n=1}^\infty (-)^n nU_n\xi^n\cr
t\tilde F_c(t)&={1\over 18}-{4\over 3}\sum_{n=1}^\infty (-)^n nV_n\xi^n
}\eqno(7.2)
$$
with the same $U_n$ and $V_n$ as in sect.2. Outside the convergence region
the ``dense polymer'' scaling functions can be found as a particular
solution to the continued equation (2.13)
$$
{1\over t}{d\over dt}t{d\over dt}\tilde U=-{1\over 2}\cosh 2\tilde U
\eqno(7.3)
$$
fixed by the initial condition at $t\to 0$
$$
\tilde U(t)=-{1\over 3}\log{t\over 8}-{1\over 2}\log\kappa+O(t^{4/3})
\eqno(7.4)
$$
Now
$$
t\tilde F_n(t)=t{d\tilde U\over dt}
\eqno(7.5)
$$
while $\tilde F_c(t)$ is restored from the continuation of eq.(2.23)
$$
t\tilde F_c(t)={1\over 18}+{1\over 2}\int_0^t\sinh 2\tilde U(r)\,rdr
\eqno(7.6)
$$
Contrary to the dilute polymer case where $U(t)$ is always regular at real
$t>0$, the continued function $\tilde U(t)$ has an infinite number of
singularities on the positive real axis. These appear as double poles of
$\exp(-2\tilde U(t))$ or as simple poles in $\tilde F_n(t)$ and $\tilde F_c
(t)$. In ref.[9] these poles were attributed to the level-crossing effect
observed in the finite-size dense polymer system. Note however that $\exp(
2\tilde U)$ has double zeroes instead of the poles and, as it is readily
figured out from eq.(7.3), remains always finite at $t>0$. Therefore $t\tilde
F_-(t)=t\tilde F_c(t)-t\tilde F_n(t)$ (the continued $tF_-(t)$ of eq.(2.32))
is positive and non-singular with an infinite sequence of critical points.

Anyhow the analytcally continued Painlev\'e III transcendent $\tilde U(t)$
worth more attention. Could one invent something as effective as the Fredholm
or TBA-like representations (3.10) and (3.21) for this case?

A closely related continuation problem arises in connection with the general
sin-Gordon scaling funciton (4.13). As one concludes from the structure (4.19)
the continuation $\eta^2\to-\eta^2$ (or equivalently $t\to e^{i\pi(p+1)/4}t$)
leads to a real expansion for the function
$$
t\tilde F_{sG}(\alpha,p|t)=tF_{sG}\left(e^{i\pi(p+1)/4}t\right)
\eqno(7.7)
$$
{}From the perturbative point of view this alternated series would correspond
to a purely imaginary coupling constant $\mu$ in the sin-Gordon action (4.1).
Being defined this way the imaginary coupled sin-Gordon model (ISG) is
apparently non-unitary and the common field theory intuition fails to figure
out the structure of its space of states, vacuum etc. However one could
proceed formally (say perturbatively) arriving at some sensible conclusions.
E.g. the renormalization group calculations in ISG with $p\gg 1$ formally work
and show that at least at $p$ large enough ISG is massless and interpolates
between two sin-Gordon critical points with different values of $p$. Moreover
the local arguments about the sin-Gordon integrability are independent on the
nature  of $\mu$. Therefore the same local higher-spin integrals of motion are
expected in the ISG as well. In ref.[9] a consistent factorized scattering
theory for the massless ISG excitations was proposed and supported by the
calculation of the ground state energy in an external field. There are some
difficulties with the TBA treatment of this scattering theory and up to now
it is not clear if it is complete or other (massive) particles are present in
the ISG spectrum.

The interest to ISG is not purely academic. As it was suggested in [9] this
field theory model is closely related to the dense phase of the 2D polymer
problem (in its scaling limit). In particular the renormalization group
behavior of the ISG effective central charge conforms qualitatively that
observed in the finite-size dense polymer system [1,9].

The TBA approach
allows the finite-size dilute polymer scaling function to be evaluated very
carefully. Unfortunately the lack of correct TBA equations impeds the same
for the dense polymer case. In this connection I'd like to mention a new
potential approach to the finite-size problem in the sin-Gordon model
developped by Destri and deVega (DdV) [19]. Contrary to TBA one does not
deal with the physical scattering theory, thermal equilibrium of physical
particles etc., but starts instead with some
``constituent particles'' and their
``bare'' scattering. After a kind of renormalization DdV arrive at a system
of integral equations which are free of the ``bare'' parameters (including
instead the ``renormalized'' physical spectra and amplitudes) and resemble
strongly the usual TBA equations. It is not yet clear if this approach can
be generalized to other integrable relativistic models or how the
Takahashi-Suzuki TBA equations are related to the DdV ones. However the DdV
system is verified (numerically) to predict the same finite-size energy as the
TBA approach does, being at the same time in many respects much more convenient
(in particular the DdV system is continuous in $p$, as opposed to the
Takahashi-Suzuki TBA). Moreover a slight modification of the DdV system looks
quite suitable for the analytic continuation (7.7) of the sin-Gordon scaling
funciton. This modified DdV system will be reported elsewhere.

Finally, the following remark seems in order. The form of potential (3.9)
suggests to consider a generalization
$$
2u(\theta)=\ldots+\bar t_3e^{-3\theta}+\bar t_1e^{-\theta}+t_1e^\theta+t_3
e^{3\theta}+t_5e^{5\theta}+\ldots+r(\theta)
\eqno(7.8)
$$
It is readily verified that in this case (in the same notations as in eq.(3.8))
$$
\eqalign{
{\partial W_1\over\partial t_3}-{\partial^3W_1\over\partial t_1^3}&=0\cr
{\partial W_3\over\partial t_3}-{\partial^3W_3\over\partial t_1^3}&=
-24\left({\partial W_1\over\partial t_1}\right)^3
}\eqno(7.9)
$$
A natural guess is that in general the function (3.10) satisfies
$$
{\partial U\over\partial t_3}={\partial^3 U\over\partial t_1^3}-
2\left({\partial U\over\partial t_1}\right)^3
\eqno(7.10)
$$
i.e. the modified KdV equation. Considering derivatives in $t_5$ etc. we can
speculate some higher integrable differential equations of the KdV hierarchy,
$U$ being related to the corresponding $\tau$-function.

\vskip 0.5cm

{\bf Acknowledgments\hfill}

Discussions with V.Fateev and H.deVega were very stimulating.

\vfill
\eject

\hbox{}
\vskip 1cm
\centerline{\bf References}
\vskip 8pt

\parskip 4pt
\parindent 35pt
\hsize=5.8in

\hskip -48pt 1. {\narrower
P.Fendley and H.Saleur. Nucl.Phys., B388 (1992) 609.
\smallskip}

\hskip -48pt 2. {\narrower
S.Cecotti, P.Fendley, K.Intrilligator and C.Vafa. Nucl.Phys., B386 (1992) 405.
\smallskip}

\hskip -48pt 3. {\narrower
T.Wu, B.McCoy, C.Tracy and E.Barouch. Phys.Rev., B13 (1976) 316.
\smallskip}

\hskip -48pt 4. {\narrower
S.Cecotti and C.Vafa. Nucl.Phys., B367 (1991) 359.
\smallskip}

\hskip -48pt 5. {\narrower
S.Cecotti. Int.J.Mod.Phys., A6 (1991) 1749; Nucl.Phys., B335 (1991) 755.
\smallskip}

\hskip -48pt 6. {\narrower
C.Vafa and N.Warner. Phys.Lett., 218B (1989) 51; W.Lerche, C.Vafa and
N.Warner. Nucl.Phys., B324 (1989) 427; E.Witten. Nucl.Phys., B340 (1990) 281;
T.Eguchi and S.-K.Yang. Mod.Phys.Lett., A4 (1990) 1653; C.Vafa. Mod.Phys.
Lett.,
A6 (1991) 337.
\smallskip}

\hskip -48pt 7. {\narrower
C.Destri and H.deVega. Nucl.Phys., B358 (1991) 251.
\smallskip}

\hskip -48pt 8. {\narrower
Al.Zamolodchikov. Mass scale in sin-Gordon and its reductions. LPM-93-06, 1993.
\smallskip}

\hskip -48pt 9. {\narrower
P.Fendley, H.Saleur and Al.Zamolodchikov. Int.J.Mod.Phys., A8 (1993) 5717;
5751.
\smallskip}

\hskip -53pt 10. {\narrower
J.Cardy and G.Mussardo. Universal properties of self-avoiding walks from
two-dimensional field theory. UCSBTH-93-12, ISAS-93-75, 1993.
\smallskip}

\hskip -53pt 11. {\narrower
B.McCoy, C.Tracy and T.T.Wu. J.Math.Phys., 18 (1977) 1058.
\smallskip}

\hskip -53pt 12. {\narrower
A.Zamolodchikov and Al.Zamolodchikov. Ann.Phys., 120 (1979) 253.
\smallskip}

\hskip -53pt 13. {\narrower
H.Bl\"ote, J.Cardy and M.Nightingale. Phys.Rev.Lett., 56 (1986) 742; I.Affleck.
Phys.Rev.Lett., 56 (1986) 746.
\smallskip}

\hskip -53pt 14. {\narrower
B.Nienhuis. J.Stat.Phys., 34 (1984) 153.
\smallskip}

\hskip -53pt 15. {\narrower
G.Waterson. Phys.Lett., 171B (1986) 77.
\smallskip}

\hskip -53pt 16. {\narrower
F.Smirnov. Phys.Lett., 275B (1992) 109.
\smallskip}

\hskip -53pt 17. {\narrower
Vl.Dotsenko and V.Fateev. Nucl.Phys., B240[FS12] (1984) 312; B251[FS13] (1985)
691.
\smallskip}

\hskip -53pt 18. {\narrower
M.Takahashi and M.Suzuki. Progr.Theor.Phys., 48 (1972) 2187.
\smallskip}

\hskip -53pt 19. {\narrower
C.Destri and H.deVega. Phys.Rev.Lett., 69 (1992) 2313.
\smallskip}

\vfill
\eject

\hbox{}
\parindent 20pt
\parskip 12pt plus 1pt
\vskip 3cm
\centerline{\bf Figure Captions}
\vskip 4pt
\hskip -50pt Fig.1.
{\narrower
Topologically different configurations of two closed polymers.
\smallskip}

\hskip -50pt Fig.2.
{\narrower
Infinite honeycomb cylinder.
\smallskip}

\hskip -50pt Fig.3.
{\narrower
Winding and non-winding polymers on a cylinder.
\smallskip}

\hskip -50pt Fig.4.
{\narrower
Distinct cylinder configurations of two closed polymer loops.
\smallskip}

\hskip -50pt Fig.5.
{\narrower
Non-contractible one-loop scaling function $tF_n(t)$.
\smallskip}

\hskip -50pt Fig.6.
{\narrower
Contractible one-loop scaling function $tF_c(t)+t^2/4$ with the bulk term
subtracted.
\smallskip}

\hskip -50pt Fig.7.
{\narrower
Scaling functions $tF_+(t)+t^2/4$ (solid curve) and $tF_-(t)+t^2/4$
(dashed one).
\smallskip}

\vfill
\eject

\end